\documentclass[11pt,a4paper]{article}
\usepackage[english]{babel}
\usepackage{amsmath,amsthm,amssymb,epsfig,latexsym}
\usepackage{color}
\usepackage{ulem}
\usepackage[numbers,square,sort&compress]{natbib}

\newcommand{\so}{\scriptscriptstyle \rm I}
\newcommand{\st}{\scriptscriptstyle \rm I\hspace{-1pt}I}
\newcommand{\sth}{\scriptscriptstyle \rm I\hspace{-1pt}I\hspace{-1pt}I}

\newcommand{\bu}{\bar u}
\newcommand{\bv}{\bar v}

\newcommand{\bet}{\bar\eta}

\newcommand{\be}[1]{\begin{equation}\label{#1}}
\newcommand{\ba}[1]{\begin{multline}\label{#1}}
\newcommand{\ee}{\end{equation}}
\newcommand{\ea}{\end{multline}}

\newcommand{\diag}{\mathop{\rm diag}}

\newtheorem{thm}{Theorem}[section]

 \makeatletter
 \@addtoreset{equation}{section}
 \makeatother
 
\newcommand{\bea}{\begin{eqnarray}}
\newcommand{\eea}{\end{eqnarray}}

\def\F{{\mathcal{F}}}
\def\E{{\mathcal{E}}}

\def\BB{{\mathbb{B}}}
\def\CC{{\mathbb{C}}}
\newcommand{\ZZ}{{\mathbb Z}}
\def\TT{{\mathbb{T}}}
\def\FF{{\rm F}}

\def\tSym{\overline{\rm Sym}}
\def\rvac{|0\rangle}

\def\EE{{\rm E}}
\def\FF{{\rm F}}

\def\TT{{T}}
\def\tEE{\tilde{\rm E}}
\def\tFF{\tilde{\rm F}}

\def\hTT{\widehat{T}}

\def\Ee{{\cal E}}

\def\sk#1{\left(#1\right)}
\def\Bal{\mathcal{B}}
\def\DBal{\mathcal{DB}}

\def\ot{\otimes}
\def\llord{\lessdot}

\def\Pep{{P}^+_e}
\def\Pem{{P}^-_e}

\def\Pfp{{P}^+_f}
\def\Pfm{{P}^-_f}
\def\oU{\overline{U}}
\def\cA{{\cal A}}
\def\cB{{\cal B}}

\def\fgo{\mathfrak{f}}
\def\ggo{\mathfrak{g}}
\def\hgo{\mathfrak{h}}
\def\hBB{\widehat{\mathbb{B}}}
\def\Dfun{\gamma}
\def\BF{\mathbb{F}}

\def\bet{\bar\eta}

\begin{document}
\pagestyle{empty}
\setcounter{page}{0}

\vspace{12pt}

\begin{center}
\begin{LARGE}
{\bf Bethe vectors for orthogonal integrable models}
\end{LARGE}

\vspace{40pt}

\begin{large}
{A.~Liashyk${}^{a}$,
S.~Z.~Pakuliak${}^{b}$,\\ E.~Ragoucy${}^c$, N.~A.~Slavnov${}^b$\  \footnote{
a.liashyk@gmail.com, stanislav.pakuliak@jinr.ru, eric.ragoucy@lapth.cnrs.fr, nslavnov@mi-ras.ru}}
\end{large}

\vspace{15mm}

${}^a$ {\it Skolkovo Institute of Science and Technology, Moscow, Russia}
\vspace{2mm}

${}^b$ {\it Steklov Mathematical Institute of Russian Academy of Sciences,\\ 8 Gubkina str., Moscow, 119991, Russia}

\vspace{2mm}

${}^c$ {\it Laboratoire de Physique Th\'eorique LAPTh, CNRS and USMB,\\
BP 110, 74941 Annecy-le-Vieux Cedex, France}

\end{center}


\vspace{4mm}

\begin{abstract}
We consider quantum integrable models associated with $\mathfrak{so}_3$ algebra.
We describe Bethe vectors of these models in terms of the current generators
of the $\mathcal{D}Y(\mathfrak{so}_3)$  algebra. To implement this approach we use  isomorphism between  $R$-matrix and Drinfeld current realizations
of the Yangians and their doubles for classical types $B$, $C$, and $D$ series algebras.
Using these results we derive the actions of the monodromy matrix elements on  off-shell
Bethe vectors. We show that these action formulas lead to recursions for off-shell Bethe vectors and Bethe equations for
on-shell Bethe vectors.  The action formulas can also be used for calculating the scalar products
in the models associated with  $\mathfrak{so}_3$ algebra.

\end{abstract}

\newpage
\pagestyle{plain}

\section{Introduction}

The algebraic Bethe ansatz \cite{FadST79,FadT79} is a powerful method to investigate quantum integrable models.
It was mostly applied to the models associated with different deformations and generalizations of the $A$-series type
algebras. The models corresponding to the $B$, $C$, and $D$ series algebras are less investigated regardless
the pioneering papers \cite{R83,R84}.

The nested algebraic Bethe ansatz \cite{KulRes81,KulRes82,KulRes83} was also developed mainly for the quantum integrable
models associated with $A$-series type algebras of the higher rank. A powerful enough approach to the
nested Bethe ansatz developed in the papers \cite{VT94,VTcom13} was recently reformulated using
the language of the current realization of the deformed infinite-dimensional algebras,  having in addition to
the current presentation, the so-called $RTT$ realization \cite{RS90} with $R$-matrices defining the fundamental
commutation relations of the monodromy matrices of the integrable models. Recently such an isomorphism between
current and $RTT$ realizations was constructed in the papers \cite{JLM18,JLY18} for the Yangians $Y(\mathfrak{so}_{2n+1})$,
$Y(\mathfrak{sp}_{2n+2})$, $Y(\mathfrak{so}_{2n+2})$, $n=1,2,\ldots,$ corresponding to the algebras of the
classical $B$, $C$, and $D$-series.
This result immediately opens a possibility to address the algebraic Bethe ansatz method to the models
with $\mathfrak{so}_{2n+1}$,  $\mathfrak{sp}_{2n+2}$, or $\mathfrak{so}_{2n+2}$ symmetries using
the current approach \cite{HLPRS17}.

In this paper we restrict ourselves to the simplest possible case of the  $\mathfrak{so}_{3}$-invariant quantum
integrable models. The corresponding $R$-matrix was found in the seminal paper by A.B.~Zamolodchikov and Al.B.~Zamolodchikov \cite{ZZ79}.
Our main concern here is calculating  the monodromy matrix elements actions on off-shell
Bethe vectors. The latter are defined within the framework of the approach introduced  in \cite{KhoPak05,KhP08}.
These action formulas turn out to be more important and more fundamental than the explicit formulas
for the Bethe vectors in terms of the monodromy matrix elements. The action of the upper-triangular elements produces
recursions for the Bethe vectors, which can be used to restore the explicit expressions for the
Bethe vectors. The action of the diagonal elements yields the Bethe equations as the condition for the off-shell Bethe
vectors to become on-shell. The action of the low-triangular monodromy matrix elements can be used for
calculation of the scalar products of the Bethe vectors \cite{BPRS12,HutLPRS17}.

The paper is organized as follows. Section~\ref{orth} gives the definition of $\mathfrak{so}_{2n+1}$ quantum integrable
model. A description of this model for $n=1$
in the language of the double Yangian $\mathcal{D}Y(\mathfrak{so}_3)$ is given in section~\ref{S-GCUM}.
Here we also introduce projections onto intersections
of the different type Borel subalgebras in this algebra and their properties.
In section~\ref{BVsect} the universal off-shell Bethe vectors are defined in terms of the current generators
of the $\mathcal{D}Y(\mathfrak{so}_3)$ algebra. This section also contains the main result of this paper describing the monodromy matrix elements actions on the
Bethe vectors.  Section~\ref{proofs} is devoted to the proof of the action formulas.

\section{Definition of the universal orthogonal integrable model}\label{orth}

In this section, we give a definition of $\mathfrak{so}_N$-invariant integrable models for $N=2n+1$.
The $\mathfrak{so}_N$-invariant $R$-matrix $R(u,v)$ has the following form \cite{ZZ79}:

\begin{equation}\label{Rmat}
  R(u,v) = \mathbf{I}\otimes\mathbf{I} + \frac{c\, \mathbf{P}}{u-v} - \frac{c\,\mathbf{Q}}{u-v+c\kappa}\,.
\end{equation}
Here $\mathbf{I}=\sum_{i=1}^N\Ee_{ii}$ is the identity operator acting in the space $\mathbf{C}^N$, $\Ee_{ij}$ are $N\times N$ matrices
with the only nonzero entry
equals to $1$ at the intersection of the $i$-th row and $j$-th column. The operators $\mathbf{P}$ and $\mathbf{Q}$ act in
$\mathbf{C}^N\otimes\mathbf{C}^N$ and respectively are given by
\begin{equation}\label{PQ}
\mathbf{P}=\sum_{i,j=1}^N\Ee_{ij}\otimes\Ee_{ji}, \qquad \mathbf{Q}=\sum_{i,j=1}^N\Ee_{ij}\otimes\Ee_{i'j'},
\end{equation}
where $i'=N+1-i$, $j'=N+1-j$. Finally, $c$ is a constant, $u$ and $v$ are arbitrary complex numbers called spectral parameters,
and $\kappa=N/2-1$.

A universal orthogonal integrable model is defined by a $N\times N$ monodromy matrix $T(u)$
whose operator-valued
entries $T_{i,j}(u)$ act in a Hilbert space $\mathcal{H}$  (the physical space of a quantum model).
We do not specify the Hilbert space $\mathcal{H}$ as well as any concrete representation of the operators $T_{i,j}(u)$: such  monodromy matrix is called universal.
It satisfies an $RTT$-algebra
\begin{equation}\label{rtt}
  R(u,v) \left( T(u)\otimes\mathbf{I} \right) \left( \mathbf{I}\otimes T(v) \right) = \left( \mathbf{I}\otimes T(v) \right) \left( T(u)\otimes\mathbf{I} \right) R(u,v).
\end{equation}
Equation \eqref{rtt} yields  commutation relations  of the monodromy matrix  entries
\begin{equation}\label{rrt2}
\begin{split}
  \left[ T_{i,j}(u), T_{k,l}(v) \right] &=  \frac{c}{u-v}\left( T_{k,j}(v)T_{i,l}(u) - T_{k,j}(u) T_{i,l}(v) \right)\\
  &+\frac{c}{u-v+c\kappa}\left(\delta_{ki'}\sum_{p=1}^N T_{p,j}(u)T_{p',l}(v)-\delta_{lj'}\sum_{p=1}^N T_{k,p'}(v)T_{i,p}(u)\right).
  \end{split}
\end{equation}

For any matrix $X$ acting in $\mathbf{C}^N$ we denote by $X^t$ the transposition
\begin{equation}\label{trans}
(X^t)_{i,j}=X_{j',i'}=X_{N+1-j,N+1-i}\,.
\end{equation}
It is related to the `usual' transposition $(\cdot)^T$ by a conjugation by the matrix $U=\sum_{i=1}^N \Ee_{ii'}$.
Note that the $R$-matrix obeys the relation
\begin{equation}\label{R-trans}
R(u,v)^{t_1t_2}=R(u,v)\,,
\end{equation}
where $t_1$ and $t_2$ stand for the transposition in the first and second spaces of $R(u,v)$ respectively.
The direct consequence of the commutation relations \eqref{rrt2}  is
an equation \cite{M07}
\begin{equation}\label{rela}
T^t(u-c\kappa)\cdot T(u) =T(u )\cdot T^t(u-c\kappa) =z(u)\mathbf{I},
\end{equation}
where $z(u)$ is a scalar commuting with all the generators $T_{i,j}(u)$.  In what follows, we set
this central element  equal to one: $z(u)=1$.

Further on we will restrict ourself to the  quantum integrable models such that
dependence of the universal monodromy matrix elements $T_{i,j}(u)$ on the parameter $u$
is given by the series
\begin{equation}\label{depen}
T_{i,j}(u)=\delta_{ij}\mathbf{1}+\sum_{\ell\geq0} T_{i,j}[\ell]u^{-\ell-1} ,
\end{equation}
where $\mathbf{1}$ and $T_{i,j}[\ell]$ respectively are the identity and nontrivial operators acting in the Hilbert space $\mathcal{H}$.
In this case, the universal monodromy matrix elements obeying \eqref{rrt2} and \eqref{rela} can be identified
with generating series of the generators of the Yangian $Y(\mathfrak{so}_N)$ and the Hilbert space $\mathcal{H}$
with the representation space of this infinite-dimensional algebra.
In particular, a direct consequence of the commutation relations \eqref{rrt2} and expansion \eqref{depen} are the
commutation relations
\begin{equation}\label{rrt2zm}
  \bigl[ T_{i,j}(u), T_{k,l}[0] \bigr] = c\sk{  \delta_{il} T_{k,j}(u) - \delta_{kj} T_{i,l}(u) - \delta_{ik'} T_{l',j}(u)+\delta_{l'j} T_{i,k'}(u)  },
\end{equation}
which will be extensively used lately. We wish to point out that the Yangian $Y(\mathfrak{so}_N)$ is defined by the relation \eqref{rrt2}
 (with the expansion \eqref{depen}) \textit{and} the coset by the relation $z(u)=1$. If one does not impose this last relation and keep $z(u)$ arbitrary (but central), one gets a bigger algebra, denoted $X(\mathfrak{o}_N)$, see e.g. \cite{AMR06}.

It follows from \eqref{rtt} that
\begin{equation*}
  \left[ \mathcal{T} (u), \mathcal{T} (v) \right]=0,
\end{equation*}
where $\mathcal{T}(u)= \sum_{i} T_{i,i}(u)$ is the universal transfer matrix. Thus, the transfer matrix is a generating function for the integrals
of motion of the model under consideration.

The key problem of the algebraic Bethe ansatz is to find eigenvectors of the universal transfer matrix $\mathcal{T}(u)$ in the space $\mathcal{H}$.
In this  context,  it is
usually assumed that the physical space of the model possesses a special reference vector $\rvac\in\mathcal{H}$ such that
\begin{equation}\label{vac}
\begin{aligned}
&T_{i,j}(u)\rvac =0,\quad i>j,\\
&T_{i,i}(u)\rvac=\lambda_i(u)\rvac,
\end{aligned}
\end{equation}
where $\lambda_i(u)$ are complex-valued functions. Then the
eigenvectors of $\mathcal{T}(u)$ are constructed as certain polynomials of the monodromy matrix entries $T_{i,j}(u)$
with $i<j$ acting on the reference vector $\rvac$. Within the framework of the universal orthogonal model associated with $R$-matrix
\eqref{Rmat}, the functions
$\lambda_i(u)$ are  free functional parameters modulo certain relations following from \eqref{rela} and which will be described
below.

We denote the algebra of the monodromy matrix elements $T_{i,j}(u)$ satisfying \eqref{rrt2} and \eqref{rela}
by $\Bal_n$ (recall that $n=(N-1)/2$). Then, the space $\mathcal{H}$ obeying \eqref{vac}
describes the whole class of $\Bal_n$ highest weight representations.

\section{Gauss coordinates of the universal monodromy\label{S-GCUM}}

In the case of $A$-series algebras, an effective way to solve eigenvalue problem for the transfer matrix is to use
instead of  the monodromy matrix elements, another set of generators associated to Gauss coordinates
of the monodromy matrix \cite{KhP08,HLPRS17}. Moreover, using recent result of \cite{JLM18,JLY18} one may verify that
the Gauss coordinates of the monodromy matrix can be used for the effective resolution of  the constraint
\eqref{rela} and  for obtaining a set of algebraically independent generators of the $RTT$
algebras related to the classical $B$, $C$, and $D$ series. On the other hand, the Gauss coordinates
relate the $RTT$ realization and the Drinfeld current presentation \cite{D88} of the quantum affine algebras and
the Yangian doubles. This allows to construct  the off-shell Bethe vectors
of the universal quantum integrable model in terms of the current generators of the corresponding
infinite dimensional algebras \cite{EKhP07}.

From now on, we restrict ourselves to the $\mathfrak{so}_3$-invariant integrable models. Thus, we deal with
the $\Bal_1$ algebra, the monodromy matrix is a $3\times 3$ matrix,  and $\kappa=1/2$ in equation \eqref{Rmat}.

Gauss coordinates for the monodromy matrix $T(u)$ can be introduced in several different
ways. In this paper we  use the following decomposition
\begin{equation}\label{Gmat}
T(u)=\mathbf{F}(u)\cdot \mathbf{D}(u)\cdot \mathbf{E}(u)\,,
\end{equation}
where $\mathbf{D}(u)$ is a diagonal  matrix
\begin{equation*}
\mathbf{D}(u)=\mbox{diag}(k_1(u),k_2(u),k_3(u)).
\end{equation*}
The matrices $\mathbf{F}(u)$ and $\mathbf{E}(u)$ respectively are
upper-triangular and lower-triangular matrices:
\begin{equation}\label{FEdef}
\mathbf{F}(u)=\begin{pmatrix} 1&\FF_{2,1}(u)&\FF_{3,1}(u)\\0&1&\FF_{3,2}(u)\\0&0&1
\end{pmatrix}, \qquad
\mathbf{E}(u)=\begin{pmatrix} 1&0&0\\\EE_{1,2}(u)&1&0\\ \EE_{1,3}(u)&\EE_{2,3}(u)&1\end{pmatrix}.
\end{equation}
Explicitly, Gauss decomposition of the monodromy  matrix associated with the  $\Bal_1$ algebra reads
\begin{equation}\label{Gauss0}
\TT(u)=\begin{pmatrix}
k_1+\FF_{2,1}k_2\EE_{1,2}+\FF_{3,1}k_3\EE_{1,3}& \FF_{2,1}k_2+\FF_{3,1}k_3\EE_{2,3}&\FF_{3,1}k_3\\
k_2\EE_{1,2}+\FF_{3,2}k_3\EE_{1,3}&k_2+\FF_{3,2}k_3\EE_{2,3}& \FF_{3,2} k_3\\
k_3\EE_{1,3}& k_3\EE_{2,3}&k_3
\end{pmatrix} ,
\end{equation}
where, for brevity, we omitted  the dependence on the
spectral parameter $u$ for all the Gauss coordinates $\EE_{i,j}(u)$, $\FF_{j,i}(u)$, and $k_i(u)$.

In terms of monodromy matrix elements, formula \eqref{Gmat} may be written as
\begin{equation}\label{Gauss}
T_{i,j}(u)=\sum_{{\rm max}(i,j)\leq\ell\leq 3} \FF_{\ell,i}(u)k_\ell(u)\EE_{j,\ell}(u)
\end{equation}
where according to \eqref{FEdef} we have $\FF_{i,i}(u)=\EE_{i,i}(u)=1$. Conditions \eqref{vac} are then ensured by the relations
$E_{j,i}(u)\rvac=0$ and $k_i(u)\rvac=\lambda_i(u)\rvac$ that we will assume from now on.

\subsection{Independent Gauss coordinates\label{SS-IGC}}

It is easy to see from  \eqref{Gauss} that assuming the invertibility of $k_i(u)$ one can express all the Gauss coordinates through  the monodromy
matrix elements $T_{i,j}(u)$.  Due to the relation \eqref{rela} these Gauss coordinates are not independent.
Let us find an independent set of the generators for the $\Bal_1$ algebra  and derive their
commutation relations.

We call an element from the  $\Bal_1$ algebra normal ordered if all the Gauss coordinate $\FF_{i,j}(u)$ are on the left and all the Gauss coordinates
$\EE_{i,j}(u)$ are on the right of this element. One can see that the Gauss decomposition
\eqref{Gauss} of the monodromy matrix is by definition normal ordered.

Due to \eqref{R-trans} the transpose-inverse monodromy matrix
\begin{equation}\label{in-tr-mo}
\hTT(u)=\Big(T(u)^{-1}\Big)^t
\end{equation}
satisfies the same $RTT$ commutation relations \eqref{rtt}.
In order to describe the  matrix $\widehat{T}(u)$ in terms of the Gauss
coordinates $\FF_{j,i}(u)$, $\EE_{i,j}(u)$, $k_i(u)$ we have to invert the matrices
$\mathbf{F}(u)$, $\mathbf{D}(u)$, and $\mathbf{E}(u)$. They are given by the relations
\begin{equation}\label{inm}
\begin{split}
\mathbf{F}(u)^{-1}&=\mathbf{I}+\textstyle{\sum_{i<j}}\Ee_{ij}\ \tFF_{j,i}(u),\\
\mathbf{D}(u)^{-1}&=\mbox{diag}(k_1(u)^{-1},k_2(u)^{-1},k_3(u)^{-1}),\\
\mathbf{E}(u)^{-1}&=\mathbf{I}+\textstyle{\sum_{i<j}}\Ee_{ji}\ \tEE_{i,j}(u),
\end{split}
\end{equation}
where $\tFF_{i,i}(u)=\tEE_{i,i}(u)=1$ and
\begin{equation}\label{tFF}
\tFF_{i+1,i}(u) =-\FF_{i+1,i}(u),\quad i=1,2,\quad  \tFF_{3,1}(u) =-\FF_{3,1}(u)+   \FF_{2,1}(u)  \FF_{3,2}(u),
\end{equation}
\begin{equation}\label{tEE}
\tEE_{i,i+1}(u) =-\EE_{i,i+1}(u),\quad i=1,2,\quad  \tEE_{1,3}(u) =-\EE_{1,3}(u)+   \EE_{2,3}(u)  \EE_{1,2}(u).
\end{equation}

The matrix elements of the transpose-inverse monodromy matrix can be easily expressed in terms of the original Gauss
coordinates $\FF_{j,i}(u)$, $\EE_{i,j}(u)$, $k_i(u)$
\begin{equation}\label{tiM}
\hTT_{i,j}(u)=\sum_{1\leq\ell\leq{\rm min}(4-i,4-j)}\tEE_{\ell,4-j}(u)k_\ell^{-1}(u)\tFF_{4-i,\ell}(u),
\end{equation}
or explicitly
\begin{equation}\label{ca4}
\hTT(u)=
\begin{pmatrix}
k_3^{-1}+\tEE_{2,3}k_2^{-1}\tFF_{3,2}+\tEE_{1,3}k_1^{-1}\tFF_{3,1}&k_2^{-1}\tFF_{3,2}+\tEE_{1,2}k_1^{-1}\tFF_{3,1} &k_1^{-1}\tFF_{3,1}\\
\tEE_{2,3}k_2^{-1}+\tEE_{1,3}k_1^{-1}\tFF_{2,1}&k_2^{-1}+\tEE_{1,2}k_1^{-1}\tFF_{2,1}&k_1^{-1}\tFF_{2,1} \\
\tEE_{1,3}k_1^{-1}&\tEE_{1,2}k_1^{-1}
&k_1^{-1}
\end{pmatrix},
\end{equation}
where as in \eqref{Gauss}, we omitted the spectral parameter dependence in the Gauss coordinates $\tFF_{j,i}(u)$, $\tEE_{i,j}(u)$, and $k_i(u)$.
In order to fix the set of algebraically independent generators of the $\Bal_1$ algebra,
we consider the relations
\begin{equation}\label{ident}
T_{i,j}(u-c/2)=\hTT_{i,j}(u),
\end{equation}
 for $2\leq i,j\leq 3$.  We have for $i,j=3$
\begin{equation}\label{k1in}
k_1(u)=k_3(u-c/2)^{-1}.
\end{equation}
To proceed further we have to normal order the Gauss coordinates in the monodromy matrix elements in \eqref{tiM}
using
\begin{equation}\label{Gcom4}
\begin{split}
k_{3}(u)^{-1}\FF_{3,2}(u)k_{3}(u)&=\FF_{3,2}(u+c),\\
k_{3}(u)\EE_{2,3}(u)k_{3}(u)^{-1}&=\EE_{2,3}(u+c),\\
[\EE_{2,3}(u),\FF_{3,2}(u-c)]&=k_2(u-c)k_3(u-c)-k_2(u)k_3(u).
\end{split}
\end{equation}
These equations are particular cases of the commutation relations
\begin{equation}\label{GLcom}
\begin{split}
k_{3}(u)\FF_{3,2}(v)k_{3}(u)^{-1}&=f(u,v)\FF_{3,2}(v)-g(u,v)\FF_{3,2}(u),\\
k_{3}(u)^{-1}\EE_{2,3}(v)k_{3}(u)&=f(u,v)\EE_{2,3}(v)-g(u,v)\EE_{2,3}(u),
\end{split}
\end{equation}
and
\begin{equation}\label{GLcomEF}
[\EE_{2,3}(u),\FF_{3,2}(v)]=g(u,v)\left(k_2(v)k_{3}(v)^{-1}-k_2(u)k_{3}(u)^{-1}\right),
\end{equation}
where
\begin{equation}\label{g-f-fun}
g(u,v)=\frac{c}{u-v},\quad f(u,v)=1+g(u,v)=\frac{u-v+c}{u-v}.
\end{equation}
Formulas \eqref{GLcom} and \eqref{GLcomEF} can be obtained from \eqref{rrt2} setting the subscripts $\{i,j,k,l\}$ equal to $\{3,3,2,3\}$, $\{3,3,3,2\}$, and $\{2,3,3,2\}$ respectively.
Note that these commutation relations are of the $\mathfrak{gl}$-type, where the second line of \eqref{rrt2}
does not contribute. Finally, using equations \eqref{ident} for $\{i,j\}=\{2,3\}$, $\{i,j\}=\{3,2\}$, and $\{i,j\}=\{2,2\}$,
we obtain
\begin{equation}\label{ind-set}
\FF_{2,1}(u)=-\FF_{3,2}(u+c/2),\quad \EE_{1,2}(u)=-\EE_{2,3}(u+c/2),
\end{equation}
and a constraint
\begin{equation}\label{cons1}
k_2(u)=k_3(u+c/2)k_3(u-c/2)^{-1}k_2(u+c/2)^{-1}.
\end{equation}

Thus, due to \eqref{k1in} and \eqref{ind-set} we can restrict ourselves to the Gauss coordinates
\begin{equation}\label{set-gen}
k_2(u),\quad k_3(u),\quad \FF_{3,2}(u),\quad \EE_{2,3}(u),
\end{equation}
and the only constraint \eqref{cons1}\footnote{Using results of the paper \cite{LPRS19} one can similarly determine a set of generators
of the $\Bal_{n}$  algebra corresponding to the classical algebra $\mathfrak{so}_{2n+1}$.}.
The latter can also be interpreted as fixing the central element
$z(u)$ \eqref{rela} expressed in terms of the diagonal Gauss coordinates as
\begin{equation}\label{center}
z(u) = k_{1}(u)k_{3}(u-c/2) =  k_{2}(u)k_{2}(u+c/2)k_{3}(u-c/2)k_{3}(u+c/2)^{-1}=1.
\end{equation}

Alternatively, we can choose the generating series
\begin{equation}\label{s-g-al}
k_1(u),\quad k_2(u),\quad \FF_{2,1}(u),\quad \EE_{1,2}(u)
\end{equation}
as a set of generators of the $\Bal_1$ algebra with a constraint
\begin{equation}\label{cons2}
k_2(u)=k_1(u+c/2)^{-1}k_1(u-c/2) k_2(u-c/2)^{-1}.
\end{equation}

Besides commutation relations \eqref{GLcom} we need also the commutation relations of the Gauss
coordinates $\FF_{3,2}(v)$ and $\EE_{2,3}(v)$ with diagonal coordinate $k_2(u)$ and  between themselves.
These commutation relations follow from \eqref{rrt2}:
\begin{equation}\label{NstFk}
k_2(u)\FF_{3,2}(v)k_2(u)^{-1}=\frac{f(u,v)}{f(u,v+c/2)}\FF_{3,2}(v)+g(u,v)\FF_{3,2}(u)+g(v,u+c/2)\FF_{3,2}(u+c/2),
\end{equation}
\begin{equation}\label{NstEk}
k_2(u)^{-1}\EE_{2,3}(v)k_2(u)=\frac{f(u,v)}{f(u,v+c/2)}\EE_{2,3}(v)+g(u,v)\EE_{2,3}(u)+g(v,u+c/2)\EE_{2,3}(u+c/2),
\end{equation}
and
\begin{equation}\label{NstFF}
\sk{u-v+\frac{c}{2}}\FF_{3,2}(u)\FF_{3,2}(v)-\sk{u-v-\frac{c}{2}}\FF_{3,2}(v)\FF_{3,2}(u)=
\frac{c}{2}\Big(\FF_{3,2}(u)^2+\FF_{3,2}(v)^2\Big),
\end{equation}
\begin{equation}\label{NstEE}
\sk{u-v-\frac{c}{2}}\EE_{2,3}(u)\EE_{2,3}(v)-\sk{u-v+\frac{c}{2}}\EE_{2,3}(v)\EE_{2,3}(u)=
-\frac{c}{2}\Big(\EE_{2,3}(u)^2+\EE_{2,3}(v)^2\Big).
\end{equation}
Getting \eqref{NstFF} and \eqref{NstEE} from the $RTT$ commutation relations \eqref{rrt2} we also obtain
\begin{equation}\label{Nstcc}
\FF_{3,1}(v)=-\frac{1}{2}\FF_{3,2}(v)^2\quad\mbox{and}\quad \EE_{1,3}(v)=-\frac{1}{2}\EE_{2,3}(v)^2.
\end{equation}
Note that because of the expansion \eqref{depen} the zero modes of the monodromy matrix elements
\begin{equation}\label{vanzm}
T_{i,i'}[0]=0
\end{equation}
vanish for $i=1,2,3$.

Note also that constraint \eqref{center} implies a relation between the eigenvalues $\lambda_i(u)$ \eqref{vac}
\begin{equation}\label{centerL}
\lambda_{1}(u)\lambda_{3}(u-c/2) =  \lambda_{2}(u)\lambda_{2}(u+c/2)\lambda_{3}(u-c/2)\lambda_{3}(u+c/2)^{-1}=1.
\end{equation}
Thus, $\lambda_i(u)$ are free functional parameters enjoying the condition \eqref{centerL}.

\subsection{Yangian double and its current realization}

In this section, we describe the construction of the Yangian double and define projections on intersections of the
different Borel subalgebras of this algebra.
This is necessary for the current realization of the off-shell Bethe vectors.

Summarizing the results of the previous subsection we conclude that the $\Bal_1$ algebra of the monodromy matrix elements
\eqref{Gmat} with
\begin{equation}\label{FYa}
\mathbf{F}(u)=\begin{pmatrix}1&-\FF_{3,2}(u+\frac{c}{2})&-\frac12\FF_{3,2}(u)^2\\[1mm]
0&1&\FF_{3,2}(u)\\[1mm]
0&0&1\end{pmatrix},\qquad\mathbf{E}(u)=\begin{pmatrix}1&0&0\\[1mm]
-\EE_{2,3}(u+\frac{c}{2})&1&0\\[1mm]
-\frac12\EE_{2,3}(u)^2&\EE_{2,3}(u)&1
\end{pmatrix},
\end{equation}
\begin{equation}\label{kYa}
\mathbf{D}(u)=\diag\left(k_3(u-\frac{c}{2})^{-1},k_2(u),k_3(u)\right),
\end{equation}
together with the constraint \eqref{cons1} and series expansion \eqref{depen}
is isomorphic to the Yangian $Y(\mathfrak{so}_3)$  \cite{Dr88}.  According to the quantum double
construction \cite{Dr88} the Yangian double \cite{KhT96}
${\cal D}Y(\mathfrak{so}_3)$   associated with the $\Bal_1$ algebra   is a Hopf algebra for a pair
 matrices $T^\pm(u)$ obeying the commutation relations
with $R$-matrix \eqref{Rmat}
\begin{equation}\label{rtt-dy}
  R(u,v) \left( T^\mu(u)\otimes\mathbf{I} \right) \left( \mathbf{I}\otimes T^\nu(v) \right) =
  \left( \mathbf{I}\otimes T^\nu(v) \right) \left( T^\mu(u)\otimes\mathbf{I} \right) R(u,v) ,
\end{equation}
where $\mu,\nu$ independently take the values $\pm$. Both matrices $T^\pm(u)$ have the Gauss
decomposition \eqref{Gmat} with the matrices \eqref{FYa}, \eqref{kYa} and constraint \eqref{cons1}. To distinguish them, we
equip the Gauss coordinates with the superscripts $\pm$.

The difference between the matrices
$T^+(u)$ and $T^-(u)$ lies in the different series expansion with respect to the spectral parameter $u$.
The matrix $T^+(u)$ is expanded over negative powers of $u$ as in \eqref{depen}. Thus,
it is identified with the universal monodromy matrix $T(u)$ \eqref{Gmat}.
In contrast, the monodromy matrix $T^-(u)$ is given by a series
 \begin{equation}\label{depen-}
T^-_{i,j}(u)=\delta_{ij}\mathbf{1}+\sum_{\ell<0} T_{i,j}[\ell]u^{-\ell-1}
\end{equation}
with respect to non-negative powers of the parameter $u$. We denote by $\DBal_1$ the algebra
generated by matrices $T^\pm(u)$ satisfying the commutation relations \eqref{rtt-dy}.

According to \cite{JLY18} one can write down the commutation relations in the double $\DBal_1$ in terms
of the formal generating series
\begin{equation}\label{DF-iso1}
F({u})=\FF^{+}_{3,2}({u})-\FF^{-}_{3,2}({u})\,,\quad
E({u})=\EE^{+}_{2,3}({u})-\EE^{-}_{2,3}({u})\,,
\end{equation}
as follows:
\begin{equation}\label{B1kF}
\begin{split}
k^{\pm}_{2}(u)F(v)k^{\pm}_{2}(u)^{-1}&=f(u,v)f(v,u+c/2)F(v),\\
k^{\pm}_3(u) F(v) k^{\pm}_3(u)^{-1}&= f(u,v)\ F(v),
\end{split}
\end{equation}
\begin{equation}\label{B1kE}
\begin{split}
k^{\pm}_{2}(u)^{-1}E(v)k^{\pm}_{2}(u)&=f(u,v)f(v,u+c/2)E(v),\\
k^{\pm}_3(u)^{-1}E(v)k^{\pm}_3(u)&=f(u,v)\ E(v),
\end{split}
\end{equation}
\begin{equation}\label{B1FF}
(u-v+c/2)\ F(u)F(v)=  (u-v-c/2)\  F(v)F(u),
\end{equation}
\begin{equation}\label{B1EE}
(u-v-c/2)\ E(u)E(v)=  (u-v+c/2)\  E(v)E(u),
\end{equation}
\begin{equation}\label{B1EF}
[E(u),F(v)]=c\ \delta(u,v)\Big(k^+_{2}(u)\cdot k^+_{3}(u)^{-1}-k^-_{2}(v)\cdot k^-_{3}(v)^{-1}\Big).
\end{equation}
Here the symbol $\delta (u,v)$ in \eqref{B1EF}  means
the additive $\delta$-function given by the formal series
\begin{equation}\label{delta}
\delta(u,v)=\frac{1}{u}\sum_{\ell\in\ZZ}\frac{v^\ell}{u^\ell}\;.
\end{equation}
The rational functions in the r.h.s. of the equations \eqref{B1kF} and \eqref{B1kE}
should be understood as power series with respect to  $v/u$ for conjugations by
$k^+_2(u)$ and $k^+_3(u)$ and as series in $u/v$ for $k^-_2(u)$ and $k^-_3(u)$.
We call the generating series $F(u)$ and $E(u)$ {\it currents} and the diagonal
Gauss coordinates $k^\pm_2(u)$ and $k^\pm_3(u)$ {\it Cartan currents}.

A coproduct in $\DBal_1$ is given by the standard formula
\begin{equation}\label{co1}
\Delta \sk{T^\pm_{i,j}(u)}=\sum_{k=1}^{3} T^\pm_{k,j}(u)\otimes T^\pm_{i,k}(u),
\end{equation}
where the monodromy matrix elements $T^+_{i,j}(u)$ and  $T^-_{i,j}(u)$ form two
Borel subalgebras each being isomorphic to $\Bal_1$.
Each of these $\Bal_1$ algebras is a natural Hopf subalgebra of $\DBal_1$.
We denote these standard Borel subalgebras by $U^\pm$.

It is well known \cite{EKhP07} that
one can associate another decomposition of the whole algebra into two dual subalgebras with the current realization of the double $\DBal_1$.
One of this {\it current} subalgebra $U_F$ is formed by the current $F(u)$ and Cartan currents
$k^+_3(u)$ and $k^+_2(u)$, while the other current subalgebra $U_E$ is formed by the current
$E(u)$ and `negative' Cartan currents  $k^-_2(u)$ and $k^-_3(u)$.
It is clear from \eqref{co1} that these new current Borel subalgebras are not  Hopf subalgebras
with respect to the coproduct \eqref{co1}. In order for the  subalgebras $U_F$ and $U_E$ to
become Hopf subalgebras in $\DBal_1$, one introduces a new, so called {\it Drinfeld coproduct} $\Delta^{(D)}$. It
is related to the original coproduct \eqref{co1} by the twisting procedure (see \cite{EKhP07} and
references therein).

For the generating series of the $\DBal_1$ algebra,  the Drinfeld coproduct
in  the current Borel subalgebra
$U_F$ ($j=2,3$) is given by
\begin{equation}\label{co3}
\Delta^{(D)}k^+_j(u)=k^+_j(u)\otimes k^+_j(u),\quad
\Delta^{(D)}F(u)=1\otimes F(u)+F(u)\otimes k^+_2(u)k^+_{3}(u)^{-1}.
\end{equation}
For the generators $k_j^-(u)$, $E(u)$ in  the dual current Borel subalgebra
$U_E$, it acts as follows:
\begin{equation}\label{co4}
\Delta^{(D)}k^-_j(u)=k^-_j(u)\otimes k^-_j(u),\quad
\Delta^{(D)}E(u)= E(u)\otimes 1+ k^-_2(u)k^-_{3}(u)^{-1}\otimes E(u).
\end{equation}

It is obvious that there are nonempty intersections of the  Borel subalgebras of different types
\begin{equation}\label{inter}
\begin{split}
U_F^-=U_F\cap U^-\,,&\qquad U_F^+=U_F\cap U^+\,,\\
U_E^-=U_E\cap U^-\,,&\qquad U_E^+=U_E\cap U^+\,,
\end{split}
\end{equation}
and these intersections are subalgebras in $\DBal_1$ \cite{EKhP07}. Furthermore,  they are coideals
with respect to the coproduct \eqref{co3}, \eqref{co4}
\begin{equation}\label{coid}
\begin{split}
\Delta^{(D)}(U_F^+)=U_F\ot U_F^+\,,&\qquad \Delta^{(D)}(U_F^-)=U_F^-\ot U_F\,,\\
\Delta^{(D)}(U_E^+)=U_E\ot U_E^+\,,&\qquad \Delta^{(D)}(U_E^-)=U_E^-\ot U_E\,.\\
\end{split}
\end{equation}

According to the general theory of the Cartan--Weyl construction  we can impose
a global ordering of the generators in $\DBal_1$. There are two different choices
for such an ordering. We denote the ordering relation by the symbol $\llord$ and introduce the cycling
ordering between elements of the subalgebras $U^\pm_F$ and $U^\pm_E$ as follows:
\begin{equation}\label{order}
\cdots \llord U_F^-\llord U_F^+\llord U_E^+\llord U_E^-\llord U_F^-\llord \cdots.
\end{equation}
Using this ordering rule we can say that arbitrary elements $\F\in U_F$ and $\E\in U_E$
are ordered if they are presented in the form
\begin{equation}\label{nor-form}
\F=\F_-\cdot \F_+\,,\quad \E=\E_+\cdot\E_-\,,
\end{equation}
where $\F_\pm\in U_F^\pm$ and $\E_\pm\in U_E^\pm$.

According to the general theory \cite{EKhP07}, one can define the projections of any ordered elements from the
subalgebras $U_F$ and $U_E$ onto subalgebras \eqref{inter} using  the formulas
\begin{equation}\label{proj1}
\begin{split}
\Pfp(\F_-\cdot\F_+)=\varepsilon(\F_-)\F_+\,,&
\quad \Pfm(\F_-\cdot \F_+)=\F_-\varepsilon(\F_+)\,,\quad \F_\pm\in U_F^\pm\,,\\
\Pep(\E_+\cdot\E_-)=\E_+\varepsilon(\E_-)\,,&
\quad \Pem(\E_+\cdot\E_-)=\varepsilon(\E_+)\E_-\,,\quad \E_\pm\in U_E^\pm\,,
\end{split}
\end{equation}
where the counit mapping $\varepsilon: \DBal_1\to \CC$ is defined by the rules
\begin{equation}\label{coun}
\varepsilon(F_i[\ell])=\varepsilon(E_i[\ell])=0\,,\quad \varepsilon(k_j[\ell])=0\,.
\end{equation}

Let $\oU_F$ be the extension of the algebra $U_F$ formed by the infinite sums of monomials
that are ordered products $\cA_{i_1}[\ell_1]\cdots \cA_{i_a}[\ell_a]$ with $\ell_1\leq\cdots\leq\ell_a$,
where $\cA_{i_l} [\ell_l]$ is either $F_{i_l}[\ell_l]$ or $k_{i_l}[\ell_l]$. Let us similarly define $\oU_E$ as
the extension of $U_E$ by  infinite sums of ordered products $\cB_{i_1}[\ell_1]\cdots \cB_{i_b}[\ell_b]$
with $\ell_1\geq\cdots\geq\ell_b$,
where $\cB_{i_l} [\ell_l]$ is either $E_{i_l}[\ell_l]$ or $k_{i_l}[\ell_l]$.
One can prove \cite{EKhP07} that
\begin{enumerate}
\item[(1)] the action of the projections \eqref{proj1} extends to the algebras $\oU_F$ and $\oU_E$ respectively;
\item[(2)] for any $\F\in\oU_F$ with $\Delta^{(D)}(\F)=\F^{(1)}\otimes\F^{(2)}$ we have
\begin{equation}\label{ordF}
\F=\Pfm\sk{\F^{(2)}}\cdot \Pfp\sk{\F^{(1)}}\,,
\end{equation}
\item[(3)] for any $\E\in\oU_E$ with $\Delta^{(D)}(\E)=\E^{(1)}\otimes\E^{(2)}$ we have
\begin{equation}\label{ordE}
\E=\Pep\sk{\E^{(1)}}\cdot \Pem\sk{\E^{(2)}}\,.
\end{equation}
\end{enumerate}

The formal definitions of the projections \eqref{proj1} are useful for proving fundamental properties of the projections
\eqref{ordF}, \eqref{ordE} onto intersections of the different types of Borel subalgebras.
In practical calculations, we will often use
a more `physical' method. For example, to calculate the projection $\Pfp$ of the product
of the currents $F(u_i)$, we replace each current  by the difference
of the Gauss coordinates $F(u_i)=\FF^+_{3,2}(u_i)-\FF^-_{3,2}(u_i)$ and then use the commutation relation
\begin{equation}\label{noF}
\sk{u-v+\frac{c}{2}}\FF^+_{3,2}(u)\FF^-_{3,2}(v)-\sk{u-v-\frac{c}{2}}\FF^-_{3,2}(v)\FF^+_{3,2}(u)=
\frac{c}{2}\Big(\FF^+_{3,2}(u)^2+\FF^-_{3,2}(v)^2\Big)
\end{equation}
to move all negative Gauss coordinates $\FF^-_{3,2}(u_i)$ to the left. Eventually, after such a normal ordering
of all the terms in the product of currents,  the action of the projection $\Pfp$ means the cancelation of all
summands having at least one `negative' Gauss coordinate $\FF^-_{3,2}(u_i)$ on the left.
The actions of the projections $\Pfm$, $\Pep$, and $\Pem$ can be defined similarly.

\section{Universal Bethe vectors for  $\Bal_1$ algebra\label{BVsect}}

A direct applications of the theory of projections lies in the construction of the universal off-shell Bethe vectors by calculating
the projections of the products of currents. For such a purpose, we identify the monodromy matrix of some
model with the generating series $T^+_{i,j}(u)$ obeying the $RTT$ relation with the corresponding
$R$-matrix. Then we define a universal off-shell  Bethe vector of this model as the projection
$\Pfp$ applied to the product of  currents corresponding to the simple roots of the underlying finite-dimensional
algebra. Since the universal monodromy matrix elements
$T^+_{i,j}(u)$ are expressed in terms of the  Gauss coordinates which are themselves related to the currents according
to the formulas \eqref{DF-iso1}, one can compute the action of the monodromy matrix elements
onto these  Bethe vectors. This leads to recurrent relations for the latter. On the other
hand, one can compute the projection of the product of currents to get the structure of
the universal Bethe vector. In all these calculations the main technical tool is the possibility to present
the product of currents in a normal ordered form using equations \eqref{ordF} or \eqref{ordE}.
In this section we implement this program in the case of the Yangian double $\DBal_1$.

\subsection{Off-shell Bethe vectors and projections}

Let us introduce rational functions
\begin{equation}\label{fun-go}
\ggo(u,v)=\frac{c/2}{u-v},\quad \fgo(u,v)=\frac{u-v+c/2}{u-v},\quad \hgo(u,v)=\frac{\fgo(u,v)}{\ggo(u,v)}=
\frac{u-v+c/2}{c/2}.
\end{equation}
They correspond to a rescaling $c\to c/2$ in the functions \eqref{g-f-fun}.
For a set of complex parameters $\bar u=\{u_1,u_2,...,u_r\}$ of cardinality $r$, we also introduce a product
\begin{equation}\label{ac98}
\Dfun(\bar u)=\prod_{i<j}^r\fgo(u_j,u_i),
\end{equation}
and a normalized ordered product of the currents
\begin{equation}\label{ac97}
\BF_r(\bar u)=\Dfun(\bar u)\F(\bar u)=\Dfun(\bar u)F(u_r)F(u_{r-1})\cdots F(u_2)F(u_1)\,.
\end{equation}
Note that due to the commutation relations \eqref{B1FF}  this normalized product
is symmetric with respect to any permutation of the parameters $u_j$.

In what follows we will consider the projection of $\BF(\bar u)$ (called the {\it pre-Bethe vector})
\begin{equation}\label{ac99}
\hBB_r(\bar u)=\Pfp\sk{\BF(\bar u)}=\Dfun(\bar u) \Pfp\sk{F(u_r)F(u_{r-1})\cdots F(u_2)F(u_1)},
\end{equation}
and the universal off-shell Bethe vector
\begin{equation}\label{ac9}
\BB_r(\bar u)=\hBB_r(\bar u)\rvac=\Dfun(\bar u) \Pfp\sk{F(u_r)F(u_{r-1})\cdots F(u_2)F(u_1)}\rvac.
\end{equation}
We call the complex variables $\bu$ in \eqref{ac99} and \eqref{ac9} {\it the Bethe parameters}.

The pre-Bethe vector $\hBB_r(\bar u)$ and Bethe vector itself\  $\BB_r(\bar u)$ are symmetric
with respect to permutations of the Bethe parameters. The term `off-shell' means that parameters $u_i$ are generic complex numbers.
If they satisfy a set of equations called Bethe equations, then the Bethe vectors become eigenvectors of the
universal transfer matrix $\mathcal{T}(u)$ and are called on-shell Bethe vectors.

In this section we calculate the projection \eqref{ac9} and obtain an expression for the Bethe vector
in terms of the Gauss coordinates $\FF^+_{3,2}(u_i)$.  This permits us to calculate the action
of monodromy matrix elements $T^+_{i,j}(z)$ \eqref{Gauss}  on Bethe vectors \eqref{ac9}. The action formulas
for $T^+_{i,j}(z)$ with $i<j$ yield recursions for the Bethe vectors in terms of the
upper-triangular matrix elements of the monodromy. The action of the diagonal elements $T^+_{i,i}(z)$
lead to the Bethe equations. Finally, the action of the lower-triangular elements  $T^+_{i,j}(z)$ for $i>j$
can be used for calculating the scalar products of Bethe vectors.  The latter are necessary tool for studying correlation functions of the  quantum
integrable model within the algebraic Bethe ansatz framework.

First of all, we  calculate the projection of the product of currents in \eqref{ac9}.
To do this we use an approach firstly implemented in \cite{KhP05}.
Let us rewrite the commutation relation \eqref{NstFF} between $\FF^\pm_{3,2}(u)$ and $\FF^-_{3,2}(v)$
in the form
\begin{equation}\label{pr4}
F(u)\FF^-_{3,2}(v)=\frac{\fgo(v,u)}{\fgo(u,v)}\FF^-_{3,2}(v)F(u)+\hgo(u,v)^{-1} X(u),
\end{equation}
where we denote by $X(u)$ the following combination of the Gauss coordinates:
$X(u)=\FF^+_{3,2}(u)^2-\FF^-_{3,2}(u)^2$. Using this commutation relation we can write
\begin{equation}\label{pr5}
\Pfp\sk{F(u_{r})\cdots F(u_{2})\FF^-_{3,2}(u_{1})}=\sum_{j=2}^r
\hgo(u_j,u_1)^{-1}\prod_{s=2}^{j-1} \frac{\fgo(u_1,u_s)}{\fgo(u_s,u_1)} \cdot Y_j,
\end{equation}
where the element $Y_j\in U^+_F$
\begin{equation*}
Y_j=\Pfp\sk{F(u_{r})\cdots F(u_{j+1})X(u_j)F(u_{j-1})\cdots F(u_{2})}
\end{equation*}
does not depend on the spectral parameter $u_1$ explicitly.

 Substituting in \eqref{pr5} $u_1=u_m$ for
$m=2,\ldots,r$ and replacing in the l.h.s. of this equation $\FF^-_{3,2}(u_{1})$ by the
difference $\FF^+_{3,2}(u_{1})-F(u_1)$ we obtain
\begin{equation}\label{pr6}
\Pfp\sk{F(u_{r})\cdots F(u_{2})}\FF^+_{3,2}(u_{m})=\sum_{j=2}^r
\hgo(u_j,u_m)^{-1}\prod_{s=2}^{j-1} \frac{\fgo(u_m,u_s)}{\fgo(u_s,u_m)} \cdot Y_j,
\end{equation}
where we used the properties of the projections \eqref{proj1} and the fact that square
of the total current $F(u)^2=0$ vanishes due
to the commutation relations \eqref{B1FF}.
We can consider \eqref{pr5} as a system of
linear equations for the unknown elements $Y_j\in U^+_F$ which can be found
as linear combinations of the elements $\Pfp\sk{F(u_{r})\cdots F(u_{2})}\FF^+_{3,2}(u_{m})$.
Solving equations \eqref{pr6} with respect to $Y_j$ and substituting them into \eqref{pr5} we obtain
\begin{equation}\label{bv6}
\Pfp\sk{F(u_{r})\cdots F(u_{1})}=\Pfp\sk{F(u_{r})\cdots F(u_{2})}
\FF^+_{3,2}(u_1;u_2,\ldots,u_r),
\end{equation}
where
\begin{equation}\label{bv7}
\FF^+_{3,2}(u_1;u_2,\ldots,u_r)=\FF^+_{3,2}(u_1)-
\sum_{j=2}^r  \hgo(u_j,u_1)^{-1}\prod_{\substack{s=2\\ s\not= j}}^r
\frac{\fgo(u_s,u_j)}{\fgo(u_s,u_1)}\FF^+_{3,2}(u_{j}).
\end{equation}
Thus, an off-shell Bethe vector $\BB_r(\bar u)$ can be
expressed as the ordered product of  linear combinations of the Gauss coordinates
acting on the reference vector
\begin{equation}\label{bv9}
\BB_r(\bar u) =\Dfun(\bar u)
\prod_{1\le j\le r}^{\longleftarrow}\FF^+_{3,2}(u_j;u_{j+1},\ldots,u_r)\rvac
\end{equation}
where the ordered product $\displaystyle{\prod^{\longleftarrow}_j} A_j$ of the non-commuting entries $A_j$ means
$A_rA_{r-1}\cdots A_2A_1$.

\subsection{Action of monodromy matrix elements on Bethe vectors}

Starting from this subsection we  use our standard shorthand notation for the products of rational functions \eqref{fun-go}
and the eigenvalues $\lambda_i(u)$ \eqref{vac}. We agree upon that if this function depends on a set of variables
(or two sets of variables), then one should take the product over this set. In  particular,
\begin{equation}\label{not}
\lambda_k(\bar u)=\prod_{u_i\in\bar u} \lambda_k(u_i),\qquad \hgo(u,\bar v)=\prod_{v_j\in\bar v}\hgo(u,v_j),\qquad
\fgo(\bar u,\bar v)=\prod_{u_i\in\bar u}\prod_{v_j\in\bar v}\fgo(u_i,v_j),
\end{equation}
and so on. We also introduce subsets $\bar u_i=\bar u\setminus\{u_i\}$ and $\bar u_{i,j}=\bar u\setminus\{u_i,u_j\}$ and extend the aforehand convention
to the products over these subsets, for instance,
\begin{equation}\label{noti}
\ggo(v_i,\bar v_i)=\prod_{\substack{v_j\in\bar v\\ v_j\ne v_i}}\ggo(v_i,v_j),\qquad
\fgo(\bar u_{i,j},\{u_i,u_j\})=\prod_{\substack{u_k\in\bar u\\ u_k\notin\{u_i,u_j\} }}\fgo(u_k,u_i)\fgo(u_k,u_j).
\end{equation}
By definition any product over the empty set is equal to 1. A double product is equal to $1$ if at least one of the sets is empty.

\begin{thm}\label{mainth}
The action of the monodromy matrix element $T_{i,j}(z)$ on  an off-shell Bethe vector $\BB_r(\bar u)$ \eqref{ac9}
gives a linear combination of off-shell Bethe vectors
\begin{equation}\label{main}
 T_{i,j}(z) \BB_r(\bu)= s(i,j)\; \lambda_3(z) \sum_{\{\bet_{\so},\bet_{\st},\bet_{\sth}\}\vdash \bet}
 \frac{\lambda_2(\bet_{\sth})}{\lambda_3(\bet_{\sth})}
 \frac{\fgo(\bet_{\so},\bet_{\st})\fgo(\bet_{\so},\bet_{\sth})  \fgo( \bet_{\st},\bet_{\sth})}{\hgo(\bet_{\so},z)\, \hgo(z+\frac{c}{2},\bet_{\sth}) }
 \BB_{r-i+j}(\bet_{\st}),
\end{equation}
where $s(i,j) = 2^{i-j + 1} (-1)^{\delta_{i1}+\delta_{j1}} $.
The sum is taken over partitions of the set $\bet=\{\bar u,z,z+c/2\}$ into several disjoint subsets
$\{\bet_{\so},\bet_{\st},\bet_{\sth}\}\vdash \bet$ with cardinalities $\#\bet_{\so}=i-1$ and
$\#\bet_{\sth}=3-j$.
\end{thm}
Proof of this theorem is given in the next section.

\subsection{Actions of  upper-triangular monodromy matrix elements\label{SS-ATME}}
The action of the upper-triangular monodromy matrix elements on the Bethe vector $\BB_r(\bu)$ are the most simple. In particular,
it follows from the restrictions on the cardinalities of the subsets $\#\bet_{\so}$ and
$\#\bet_{\sth}$ that $\bet_{\so}=\bet_{\sth}=\emptyset$ for the action of $T_{1,3}(z)$. The sum over partitions in \eqref{main}
disappears and we immediately arrive at
\begin{equation}\label{lac4}
T_{1,3}(z)\BB_r(\bu)=-\frac{\lambda_3(z)}{2} \BB_{r+2}\sk{\bu,z,z+\frac{c}2}.
\end{equation}

For the action of $T_{1,2}(z)$, one has $\#\bet_{\so}=0$ and $\#\bet_{\sth}=1$. The sum over partitions in \eqref{main} turns into
\begin{equation}\label{T12Act}
 T_{1,2}(z) \BB_r(\bu)= - \lambda_3(z) \sum_{\{\bet_{\st},\bet_{\sth}\}\vdash \bet}
 \frac{\lambda_2(\bet_{\sth})}{\lambda_3(\bet_{\sth})}
 \frac{\fgo( \bet_{\st},\bet_{\sth})}{\hgo(z+\frac{c}{2},\bet_{\sth}) }
 \BB_{r+1}(\bet_{\st}),
\end{equation}
where either $\bet_{\sth}=z$, or $\bet_{\sth}=z+c/2$, or $\bet_{\sth}=u_i$ with $i=1,\dots,r$. It is easy to see that the case $\bet_{\sth}=z+c/2$ does not
contribute due to $\fgo(z,z+\frac{c}2)=0$. Thus, we obtain
\begin{equation}\label{lac5}
\begin{split}
&T_{1,2}(z)\BB_r(\bu)=-\lambda_2(z)\fgo(\bu,z) \BB_{r+1}\sk{\bu,z+\frac{c}2}\\
&\qquad -\lambda_3(z)
\sum_{i=1}^r \ggo(z,u_i) \fgo(\bu_i,u_i) \frac{\lambda_2(u_i)}{\lambda_3(u_i)} \BB_{r+1}\sk{\bu_{i},z,z+\frac{c}{2}}.
\end{split}
\end{equation}
Similarly, we derive the action of $T_{2,3}(z)$:
\begin{equation}\label{lac6}
\begin{split}
&T_{2,3}(z)\BB_r(\bu)=\lambda_3(z)\fgo\sk{z+\frac{c}2,\bu} \BB_{r+1}\sk{\bu,z}\\
&\qquad -\lambda_3(z)
\sum_{i=1}^r \frac{\fgo(u_i,\bu_i)}{\hgo(z,u_i)}\BB_{r+1}\sk{\bu_{i},z,z+\frac{c}{2}}.
\end{split}
\end{equation}

Equations \eqref{lac4}--\eqref{lac6} yield recursions for the Bethe vectors
\begin{equation}\label{rec12}
\BB_{r+1}\sk{\bu,z+\frac{c}2}=-\frac{T_{1,2}(z)\BB_r(\bu)}{\lambda_2(z)\,\fgo(\bu,z)}-
\frac{2T_{1,3}(z)}{\lambda_2(z)\,\fgo(\bu,z)}\sum_{i=1}^r \frac{\fgo(\bu_i,u_i)}{\hgo(u_i,z+\frac{c}2)} \frac{\lambda_2(u_i)}{\lambda_3(u_i)}\BB_{r-1}(\bu_i),
\end{equation}
and
\begin{equation}\label{rec23}
\BB_{r+1}\sk{\bu,z}=\frac{T_{2,3}(z)\BB_r(\bu)}{\lambda_3(z)\,\fgo(z+\frac{c}2,\bu)}-
\frac{2T_{1,3}(z)}{\lambda_3(z)\,\fgo(z+\frac{c}2,\bu)}\sum_{i=1}^r \frac{\fgo(u_i,\bu_i)}{\hgo(z,u_i)}\BB_{r-1}(\bu_i).
\end{equation}
These recursions allow us to successively construct Bethe vectors in terms of polynomials in $T_{i,j}$ with $i<j$ applied to $\rvac$ starting from the initial condition
$\BB_{0}(\emptyset)=\rvac$ and $\BB_{1}(u)=T_{2,3}(u)\rvac/\lambda_3(u)$.
We failed to solve these recursions and obtain explicit compact expressions for the generic Bethe vectors in terms of such polynomials.
However, in most problems involving the algebraic Bethe ansatz, such explicit formulas are not required.

\subsection{Actions of diagonal monodromy matrix elements\label{SS-ADME}}

The actions of the diagonal operators can be derived in the same manner. We omit the details of this derivation
and show how the action formula \eqref{main} implies the Bethe equations in the case of the action of diagonal elements $T_{i,i}(z)$.

Setting
$i=j=1$ in \eqref{main} we obtain
\begin{equation}\label{lac1}
\begin{split}
&T_{1,1}(z)\BB_r(\bu) =  \lambda_1(z) \fgo(\bu,z)\fgo\big(\bu,z+\frac{c}{2}\big)\, \BB_r(\bu) \\
&\quad + 2\lambda_2(z) \sum_{i=1}^r  \ggo\big(z+\frac{c}{2},u_i\big)\frac{\lambda_2(u_i)}{\lambda_3(u_i)}
\fgo(\bu_i,z)  \fgo(\bu_i,u_i)\, \BB_r\sk{\bu_i,z+\frac{c}{2}}\\
&\quad + 2\lambda_3(z) \sum_{i<j}^r \ggo\big(z,\{u_i,u_j\}\big) \frac{\lambda_2(u_i)}{\lambda_3(u_i)} \frac{\lambda_2(u_j)}{\lambda_3(u_j)}
\fgo\big(\bu_{i,j},\{u_i,u_j\}\big)\,\BB_r\sk{\bu_{i,j},z,z+\frac{c}{2}}.
\end{split}
\end{equation}
Here the first term in r.h.s. corresponds to $\bet_{\sth}=\{z,z+\frac{c}2\}$, the second term corresponds to
$\bet_{\sth}=\{u_i,z+\frac{c}2\}$, and the third term corresponds to $\bet_{\sth}=\{u_i,u_j\}$, $i<j$. The term
corresponding to the subset $\bet_{\sth}=\{u_i,z\}$ vanishes due to $\fgo(z,z+\frac{c}2)=0$. To obtain the first term in the
r.h.s. of \eqref{lac1} we also used  \eqref{center}.

The action of $T_{2,2}(z)$ on the Bethe vector is
\begin{equation}\label{lac2}
\begin{split}
&T_{2,2}(z)\BB_r(\bu) =  \lambda_2(z) \fgo(\bu,z)\fgo\big(z+\frac{c}{2},\bu\big)\, \BB_r(\bu) \\
&\quad+ 2\lambda_3(z) \sum_{i=1}^r \ggo(z,u_i) \frac{\lambda_2(u_i)}{\lambda_3(u_i)}  \fgo\big(z+\frac{c}{2},\bu_i\big) \fgo(\bu_i,u_i)\, \BB_r(\bu_i,z)\\
&\quad - 2\lambda_2(z) \sum_{i=1}^r  \frac{\fgo(\bu_i,z)}{\hgo(z,u_i)}   \fgo(u_i,\bu_i)\, \BB_r\sk{\bu_i,z+\frac{c}{2}}\\
&\quad + 2\lambda_3(z) \sum_{ i\ne j}^r  \frac{\lambda_2(u_i)}{\lambda_3(u_i)} \frac{\lambda_2(u_j)}{\lambda_3(u_j)}
\frac{ \fgo(u_i,\bu_{i,j}) \fgo(\bu_{j},u_j)}{\hgo(z,u_i) \hgo(u_j,z+\frac{c}{2})}\, \BB_r\sk{\bu_{i,j},z,z+\frac{c}{2}}.
\end{split}
\end{equation}
Here the terms in the r.h.s. corresponds to the partitions such that $\{\bet_{\so},\bet_{\sth}\}$ are either $\{u_i,u_j\}$, or $\{u_i,z\}$, or $\{z+\frac{c}2,u_i\}$, or $\{z+\frac{c}2,z\}$. Contributions of other partitions vanish due to $\fgo(z,z+\frac{c}2)=0$.

The action of $T_{3,3}(z)$ is given by
\begin{equation}\label{lac3}
\begin{split}
&T_{3,3}(z)\BB_r(\bu) =  \lambda_3 (z)  \fgo(z,\bu)\fgo\big(z+\frac{c}{2},\bu\big)\, \BB_r(\bu)\\
&\quad + 2 \lambda_3 (z) \sum_{i=1}^r \ggo(u_i,z)  \fgo\big(z+\frac{c}{2},\bu_i\big) \fgo(u_i,\bu_i)\, \BB_r(\bu_i,z) \\
&\quad + 2 \lambda_3 (z) \sum_{i<j}^r  \frac{\fgo(\{u_i,u_j\},\bu_{i,j})}{\hgo(z,\{u_i,u_j\})}\, \BB_r\sk{\bu_{i,j},z,z+\frac{c}{2}}.
\end{split}
\end{equation}

Summing up all three equations \eqref{lac1}--\eqref{lac3} and gathering the coefficients of the Bethe vectors
$\BB_r(\bu)$, $\BB_r(\bu_i,z)$, $ \BB_r\sk{\bu_i,z+\frac{c}{2}}$, and $\BB_r\sk{\bu_{i,j},z,z+\frac{c}{2}}$ we
see that if Bethe parameters satisfy the system of Bethe equations
\begin{equation}\label{BE}
\frac{\lambda_2(u_i)}{\lambda_3(u_i)} =\frac{\fgo(u_i,\bu_i)}{\fgo(\bu_i,u_i)},
\end{equation}
then the Bethe vector $\BB_r(\bu)$ becomes the eigenvector of the transfer matrix $\mathcal{T}(z)$
\begin{equation}\label{eigen}
\mathcal{T}(z)\ \BB_r(\bu)=\tau(z|\bu)\ \BB_r(\bu),
\end{equation}
with an eigenvalue
\begin{equation}\label{eigval}
 \tau(z | \bu) = \lambda_1(z) \fgo(\bu,z)\fgo\sk{\bu,z+\frac{c}{2}} + \lambda_2(z) \fgo(\bu,z)\fgo\sk{z+\frac{c}{2},\bu} +
 \lambda_3(z) \fgo(z,\bu)\fgo\sk{z+\frac{c}{2},\bu}.
\end{equation}

Observe that due to \eqref{centerL} we can also write down the Bethe equations in the form
\begin{equation}\label{BE1}
\frac{\lambda_1(u_i-c/2)}{\lambda_2(u_i-c/2)} =\frac{\fgo(u_i,\bu_i)}{\fgo(\bu_i,u_i)}.
\end{equation}
It is easy to see that the systems \eqref{BE}, \eqref{BE1} are equivalent to the absence of the poles of the transfer matrix eigenvalue \eqref{eigval}
at $z=u_i$ and $z=u_i-c/2$.

We do not give explicit formulas for the action of lower-triangular elements of $T(u)$, since they are quite cumbersome. We only note that these formulas can be
used to compute scalar products of Bethe vectors. We get in this way
\begin{equation}\label{SP-corr}
 S_r^{\mathfrak{so}_3}(\bu|\bv) = 2^{-r}\left. S_r^{\mathfrak{gl}_2}(\bu|\bv)\right|_{c\to c/2},
\end{equation}
where $S_r^{\mathfrak{so}_3}(\bu|\bv)$ denotes the scalar product of $\BB_r(\bv)$ with the dual vector of $\BB_r(\bu)$, and $S_r^{\mathfrak{gl}_2}(\bu|\bv)$ is the same object for the generalized model based on $Y(\mathfrak{gl}_2)$.  Equation \eqref{SP-corr} is not surprising since the Yangian
$Y(\mathfrak{so}_3)$ and the Yangian $Y(\mathfrak{gl}_2)$ are isomorphic. In fact, the equality can be viewed already at the level of Bethe vectors:
\begin{equation}
 \BB_r^{\mathfrak{so}_3}(\bu) = \left. 2^{-\frac{r}2}\; \BB_r^{\mathfrak{gl}_2}(\bu)\right|_{c\to c/2}.
\end{equation}
Note that the isomorphism is rather explicit and simple in the current presentation, but more involved in terms of the monodromy matrix elements (see \cite{AMR06} for an explicit construction in the $RTT$ presentation).

\section{Proofs}\label{proofs}

In order to prove the statement of theorem~\ref{mainth} we need a special presentation for the pre-Bethe
vectors \eqref{ac99} in terms of the normalized product of currents \eqref{ac97} and the `negative' Gauss coordinates
$\FF^-_{3,2}(u_i)$. This presentation is a direct consequence of the projection properties
\eqref{ordF}. In order to formulate it, we introduce several additional notions.

For any formal series $G(\bar u)$ depending on a set of parameters $\bu=\{u_1,\ldots,u_r\}$,
 we define a deformed symmetrization by the sum
\begin{equation}\label{sy1}
\tSym_{\bar u}\ G(\bar u)=\sum _{\sigma\in S_{r}} \prod_{\substack{\ell<\ell'\\ \sigma(\ell)>\sigma(\ell')}}
\frac{\fgo(u_{\sigma(\ell')},u_{\sigma(\ell)})}{\fgo(u_{\sigma(\ell)},u_{\sigma(\ell')})}\ G(^\sigma\bar u),
\end{equation}
where $S_{r}$ is a permutation group of the set  $\bu$ and
$^\sigma\bar u=\{u_{\sigma(1)},\ldots,u_{\sigma(r)}\}$.

Let $\F(\bar u)$ be the ordered product of the currents $F(u_i)$ defined by \eqref{ac97}. $\F(\bar u)\in\oU_F$ is a formal
series of the generators of the  $\DBal_1$ algebra. It can be presented in the normal ordered
form via \eqref{ordF}, the coproduct properties \eqref{co3}, and the
commutation relations \eqref{B1kF}
\begin{equation}\label{sy2}
\F(\bar u)=\tSym_{\bar u}\sk{\sum_{s=0}^r \frac{1}{s!(r-s)!}\
\Pfm\sk{F_2(u_r)\cdots F_2(u_{s+1})}\cdot \Pfp\sk{F_2(u_s)\cdots F_2(u_{1})}}.
\end{equation}
Multiplying both sides  by the factor $\Dfun(\bu)$ \eqref{ac98} and using the fact that
for any formal series $G(\bar u)$
\begin{equation}\label{sy4}
\Dfun(\bar u)\tSym_{\bar u}\sk{G(\bar u)}=\mbox{Sym}_{\bar u} \sk{\Dfun(\bar u)G(\bar u)},
\end{equation}
we obtain the ordering rule for the normalized symmetric product of the currents
\begin{equation}\label{sy22}
\BF_r(\bar u)=\mbox{Sym}_{\bar u}\sk{\sum_{s=0}^r
\frac{\prod_{j=s+1}^r\prod_{i=1}^s \fgo(u_j,u_i)}{s!(r-s)!}\
\Pfm\bigl(\BF_{r-s}(u_r,\ldots,u_{s+1})\bigr)\cdot \Pfp\bigl(\BF_s(u_s,\ldots,u_{1})\bigr)}.
\end{equation}
Since both $\Pfm\bigl(\BF_{r-s}(u_r,\ldots,u_{s+1})\bigr)$ and $\Pfp\bigl(\BF_s(u_s,\ldots,u_{1})\bigr)$ are symmetric
over their arguments, the sum over permutations within  subsets $\{u_r,\ldots,u_{s+1}\}$ and $\{u_s,\ldots,u_{1}\}$ leads to the
cancellation of the combinatorial factor $s!(r-s)!$. The sum over permutations of the whole set $\bu$ thus turns into the sum over
partitions of this set  into two nonintersecting
subsets $\bar u_{\so}$ and $\bar u_{\st}$ with cardinalities
cardinalities
$\#\bar u_{\so}=s$ and $\#\bar u_{\st}=r-s$ for any $s$
\begin{equation}\label{sy23}
\BF_r(\bar u)= \sum_{\{\bu_{\so},\bu_{\st}\}\vdash \bu} \fgo(\bu_{\st},\bu_{\so})
\Pfm\big(\BF_{r-s}(\bu_{\st})\big)\cdot \Pfp\big(\BF_s(\bu_{\so})\big).
\end{equation}

Using the relation  $\Pfm\sk{F_2(u)}=-\FF^-_{3,2}(u)$ and calculating  the projection via \eqref{noF}
\begin{equation*}
\Pfm\sk{\BF_2(\{u_j,u_i\})}= \sk{\fgo(u_j,u_i)\FF^-_{3,2}(u_j)-\ggo(u_j,u_i)\FF^-_{3,2}(u_i)}\FF^-_{3,2}(u_i),
\end{equation*}
we can write down the pre-Bethe vector $\Pfp\sk{\BF_r(\bu)}$ in the form
\begin{equation}\label{sy6}
\begin{split}
&\Pfp\sk{\BF_r(\bar u)}=\BF_r(\bar u)+\sum_{i=1}^r\fgo(u_i,\bar u_i)\FF^-_{3,2}(u_i)\Pfp\sk{\BF_{r-1}(\bar u_i)}\\
&-\sum_{i<j}^r \fgo(\{u_j,u_i\},\bar u_{i,j})\sk{\fgo(u_j,u_i)\FF^-_{3,2}(u_j)-
\ggo(u_j,u_i)\FF^-_{3,2}(u_i)}\FF^-_{3,2}(u_i)\Pfp\sk{\BF_{r-2}(\bar u_{i,j})}+\mathbb{W}.
\end{split}
\end{equation}
Here $\mathbb{W}$ stays for all the terms which contain  at least three `negative'
Gauss coordinates $\FF^-_{3,2}(u_i)$ on the left of the product.

\subsection{Action of the elements $T_{i,3}(z)$}

Formula \eqref{bv6} for $r=2$, $u_2=u_1+\frac{c}2$ and $u_1=z$ reads
\begin{equation}\label{ac7}
\Pfp\sk{F\sk{z+\frac{c}{2}}F\sk{z}}=\FF^+_{3,2}\sk{z+\frac{c}{2}}\sk{\FF^+_{3,2}\sk{z}-\frac12\FF^+_{3,2}\sk{z+\frac{c}{2}}}.
\end{equation}
According to the commutation relation \eqref{NstFF} taken at $u=z+c/2$ and $v=z$
this projection is equal to
\begin{equation}\label{ac8}
\Pfp\sk{F\sk{z+\frac{c}{2}}F\sk{z}}=\frac12\FF^+_{3,2}\sk{z}^2.
\end{equation}
This means that the monodromy matrix element $T^+_{1,3}(z)$ can be expressed through the current generators as follows:
\begin{equation}\label{i3c3}
T^+_{1,3}(z)=-\Pfp\sk{F\sk{z+\frac{c}{2}}F\sk{z}k^+_3(z)}.
\end{equation}

On the other hand, the properties of the projection onto $U^+_F$ imply that for any $\F_1,\F_2\in \oU_F$
\begin{equation}\label{prpr}
\Pfp\sk{\F_1\cdot \Pfp\sk{\F_2}}= \Pfp\sk{\F_1}\cdot \Pfp\sk{\F_2}.
\end{equation}
The commutation relations \eqref{rrt2} in the $\DBal_1$  algebra
\begin{equation}\label{i3c1}
f(u,z)T^+_{1,3}(z)T^-_{2,3}(u)=T^-_{2,3}(u)T^+_{1,3}(z)+g(u,z)T^-_{1,3}(u)T^+_{2,3}(z),
\end{equation}
and property \eqref{prpr} yield 
\begin{equation}\label{i3c2}
\Pfp\sk{T^+_{1,3}(z)\cdot \FF^-_{3,2}(u)}=0,
\end{equation}
or  equivalently
\begin{equation}\label{i3c5}
\Pfp\sk{F\sk{z+\frac{c}{2}}F\sk{z}k^+_3(z)\cdot \FF^-_{3,2}(u)}=0.
\end{equation}

Then the action  of $T^+_{1,3}(z)$ on Bethe vector reads
\begin{equation}\label{i3c4}
\begin{split}
T^+_{1,3}(z)\cdot \BB_r(\bu)&=-\Pfp\sk{F\sk{z+\frac{c}{2}}F\sk{z}k^+_3(z)}\cdot \Pfp\sk{\BF_r(\bu)}\rvac\\
& = -\Pfp\sk{F\sk{z+\frac{c}{2}}F\sk{z}k^+_3(z)\cdot \Pfp\sk{\BF_r(\bu)}}\rvac\\
&= - \lambda_3(z)f(z,\bu)\Pfp\sk{F\sk{z+\frac{c}{2}}F\sk{z} \BF_r(\bu)}\rvac\\
&= -\frac{\lambda_3(z)}{2} \BB_{r+2}\sk{z+\frac{c}2,z,\bu}.
\end{split}
\end{equation}
Here we used \eqref{sy6} and \eqref{i3c5} to replace $ \Pfp\sk{\BF(\bu)}$ by $ \BF(\bu)$ in the second line of \eqref{i3c4}, and
the obvious relations
$\fgo\sk{z+\frac{c}2,z}=2$ and $\fgo\sk{z+\frac{c}2,u}\fgo(z,u)=f(z,u)$. Thus, the action formula \eqref{lac4} is proved.

Combining formulas \eqref{rrt2} for the values of the indices $\{i,j,k,l\}\to\{2,3,2,3\}$ and $\{3,3,1,3\}$ in the double Yangian $\DBal_1$
we  obtain  commutation relations
\begin{equation}\label{i3c6}
\begin{split}
&T^+_{2,3}(z)T^-_{2,3}(u)=\frac{\fgo\sk{z+\frac{c}2,u}}{\fgo\sk{u+\frac{c}2,z}}T^-_{2,3}(u)T^+_{2,3}(z)\\
&\qquad +\frac2{\hgo(z,u)}\sk{T^+_{1,3}(z)T^-_{3,3}(u)-\frac{\fgo\sk{z+\frac{c}2,u}\fgo\sk{z,u}}{\fgo\sk{u+\frac{c}2,z}\fgo\sk{u,z}}
T^-_{1,3}(u)T^+_{3,3}(z)}.
\end{split}
\end{equation}
Using explicit expressions for the monodromy matrix elements in terms of the Gauss coordinates we conclude from \eqref{i3c6} that
\begin{equation}\label{i3c7}
\Pfp\sk{T^+_{2,3}(z )\ \FF^-_{3,2}(u)}=\frac2{\hgo(z,u)}\ T^+_{1,3}(z).
\end{equation}
Combining this formula with equation \eqref{i3c2} we obtain
\begin{equation}\label{i3c8}
\Pfp\sk{T^+_{2,3}(z )\ \FF^-_{3,2}(u) \FF^-_{3,2}(v)}=0.
\end{equation}

Now equations \eqref{i3c7} and \eqref{i3c8} together with the presentation \eqref{sy6} yield
\begin{equation}\label{i3c9}
\begin{split}
&T^+_{2,3}(z)\cdot \BB_r(\bu)=\Pfp\sk{F(z)k^+_3(z)}\cdot \Pfp\sk{\BF_r(\bu)}\rvac\\
&\quad=\lambda_3(z)f(z,\bu)\Pfp\sk{F(z)\BF_r(\bar u)}\rvac+2\sum_{i=1}^r
\frac{\fgo(u_i,\bar u_i)}{\hgo(z,u_i)} \Pfp\sk{T^+_{1,3}(z)\Pfp\sk{\BF_{r-1}(\bar u_i)}}\rvac\\
&\quad=\lambda_3(z)\sk{\fgo\sk{z+\frac{c}2,\bu} \BB_{r+1}(\{z,\bu\} - \sum_{i=1}^r \frac{\fgo(u_i,\bar u_i)}{\hgo(z,u_i)}
\BB_{r+1}\sk{\left\{z+\frac{c}2,z,\bu\right\}}}\rvac,
\end{split}
\end{equation}
thus proving \eqref{lac6}.

Taking into account that $T^+_{3,3}(z)=k^+_3(z)$, the commutation relations \eqref{rrt2} in the double Yangian $\DBal_1$ yield
\begin{equation}\label{i3c10}
T^+_{3,3}(z)\FF^-_{3,2}(u)=f(z,u)\FF^-_{3,2}(u)T^+_{3,3}(z)+g(u,z)T^+_{2,3}(z).
\end{equation}
This formula together with \eqref{i3c2}, \eqref{i3c7}, and \eqref{i3c8} gives the following relations:
\begin{equation}\label{i3c11}
\Pfp\sk{T^+_{3,3}(z )\ \FF^-_{3,2}(u)}=g(u,z)T^+_{2,3}(z),
\end{equation}
\begin{equation}\label{i3c12}
\Pfp\sk{T^+_{3,3}(z )\ \FF^-_{3,2}(u) \FF^-_{3,2}(v)}=\frac{2g(u,z)}{\hgo(z,v)} T^+_{1,3}(z),
\end{equation}
and
\begin{equation}\label{i3c13}
\Pfp\sk{T^+_{3,3}(z )\ \FF^-_{3,2}(u) \FF^-_{3,2}(v) \FF^-_{3,2}(w)}=0.
\end{equation}

Using these equations and the first three terms in the r.h.s. of \eqref{sy6} we prove \eqref{lac3}.

One can prove all the other action formulas of theorem~\ref{mainth} in a similar way.
It is clear, however, that for calculating the action of the monodromy matrix low-triangular elements,
we should compute more terms in the  r.h.s. of the presentation \eqref{sy6}. This makes the calculations
rather cumbersome. Instead, there exists a more elegant way to calculate these actions using the zero modes of the
monodromy matrix and the commutation relations \eqref{rrt2zm}.  We explain this approach in the next subsection.

\subsection{Zero modes action}

In what follows, we use the zero modes of the low-triangular monodromy matrix elements
\begin{equation}\label{zm1}
T^+_{2,1}[0]=-T^+_{3,2}[0]=-\EE^+_{2,3}[0],
\end{equation}
and the action of the zero mode $\EE^+_{2,3}[0]$ on the Bethe vector $\BB_r(\bu)$.
To calculate this action, we use again the presentation \eqref{sy6} and the commutation relations
which follow from  \eqref{GLcomEF} and \eqref{B1kF}
\begin{equation}\label{zm2}
\EE^+_{2,3}[0]\FF^-_{3,2}(u)= \FF^-_{3,2}(u)\EE^+_{2,3}[0]+c\big(k^-_2(u)k^-_3(u)^{-1}-1\big),
\end{equation}
\begin{equation}\label{zm3}
\EE^+_{2,3}[0] F(u)=  F(u)\EE^+_{2,3}[0]+c\big(k^+_2(u)k^+_3(u)^{-1}-k^-_2(u)k^-_3(u)^{-1}\big),
\end{equation}
\begin{equation}\label{zm4}
k^+_2(u)k^+_3(u)^{-1} F(v)=  \frac{\fgo(v,u)}{\fgo(u,v)}F(v) k^+_2(u)k^+_3(u)^{-1}.
\end{equation}
Equation \eqref{zm2} means that only two terms in the r.h.s. of \eqref{sy6} contribute to the
action of the zero mode $\EE^+_{2,3}[0]$ on the Bethe vector. Since $\EE^+_{2,3}[0]\rvac=0$,
this action follows from the chain of equations
\begin{equation}\label{zm5}
\begin{split}
\EE^+_{2,3}[0]\ \BB_r(\bu)&=\left[\EE^+_{2,3}[0], \Pfp\sk{\BF_r(\bu)}\right]\rvac\\
&=\Pfp\sk{\left[\EE^+_{2,3}[0],\sk{\BF_r(\bu)-\sum_{i=1}^r\fgo(u_i,\bu_i)\FF^-_{3,2}(u_i)\BF_{r-1}(\bu_i)}\right]}\rvac\\
&=c\Pfp\sk{\sum_{i=1}^r\fgo(u_i,\bu_i)\sk{k^+_2(u_i)k^+_3(u_i)^{-1} - 1 } \BF_{r-1}(\bu_i)}\rvac\\
&=c\sum_{i=1}^r \fgo(u_i,\bu_i) \sk{\frac{\fgo(\bu_i,u_i)}{\fgo(u_i,\bu_i)}\frac{\lambda_2(u_i)}{\lambda_3(u_i)} -1}
\BB_{r-1}(\bu_i).
\end{split}
\end{equation}
Note that if the set $\bu$ satisfies the Bethe equations \eqref{BE}, then the zero mode
$\EE^+_{2,3}[0]$ annihilate the on-shell Bethe vectors, which appear to be highest weight vectors of the
finite dimensional $\mathfrak{so}_3$ algebra generated by the zero modes.
This a typical property of the on-shell
Bethe vectors  in the models associated to Yangians.
Indeed, these models are invariant under the action of the finite dimensional algebra generated by the zero modes,
and the highest weight property is related to the completeness of the Bethe ansatz.
This was already noticed for models based on $\mathfrak{gl}_N$ symmetry \cite{MukTV}.

Formula \eqref{rrt2zm} at $\{i,j,k,l\}=\{1,3,3,2\}$ yields
\begin{equation}\label{zm6}
T^+_{1,2}(z)=-T^+_{2,3}(z)+c^{-1}[\EE^+_{2,3}[0],T^+_{1,3}(z)].
\end{equation}
It can be used to obtain the action the monodromy matrix element $T^+_{1,2}(z)$
from the already known actions of  $T^+_{2,3}(z)$, $T^+_{1,3}(z)$, and $\EE^+_{2,3}[0]$.
To prove the rest of theorem~\ref{mainth},
we should use the following relations:
\begin{equation}\label{zm7}
\begin{aligned}
&T^+_{2,2}(z)=-T^+_{3,3}(z)+c^{-1}[\EE^+_{2,3}[0],T^+_{2,3}(z)],\\
&T^+_{1,1}(z)=T^+_{2,2}(z)-c^{-1}[\EE^+_{2,3}[0],T^+_{1,2}(z)],\\
&T^+_{3,2}(z)=c^{-1}[\EE^+_{2,3}[0],T^+_{3,3}(z)],\\
&T^+_{2,1}(z)=c^{-1}[\EE^+_{2,3}[0],T^+_{1,1}(z)],\\
&%
T^+_{3,1}(z)=-c^{-1}[\EE^+_{2,3}[0],T^+_{3,2}(z)].
\end{aligned}
\end{equation}
They imply the action formulas
 for $T^+_{2,2}(z)$, $T^+_{1,1}(z)$, $T^+_{3,2}(z)$, $T^+_{2,1}(z)$, and $T^+_{3,1}(z)$.

\section*{Conclusion}

This paper starts a program of investigating the $\mathfrak{so}_N$-invariant quantum integrable models
 using current formulation of the deformed Kac--Moody algebras and method of projections onto
 intersections of different type Borel subalgebras in these algebras. This method was formulated in \cite{EKhP07,KhP08} and developed in \cite{HLPRS17}.
Within the framework of this method, the off-shell Bethe vectors
are defined in terms of the generators of the corresponding infinite dimensional algebra. The $RTT$ formulation
of this algebra uses the same $R$-matrix as the intertwining relations of the monodromy matrix of
$\mathfrak{so}_n$-invariant quantum integrable model.   The main goal of the present paper was to find formulas for the action of
the monodromy matrix elements on Bethe vectors using current approach and method of projections. We have shown that our approach
allows one to obtain such  formulas and to express the result of these actions  as linear combinations of
off-shell Bethe vectors.

Note that we do not use explicit presentations for the Bethe vectors in terms of the monodromy matrix elements acting on the reference
vector. Such the explicit representations are missing up to now, though the recursions derived in this paper allow one to find
them at least for the vectors with small number of Bethe parameters. We have shown that the projection method allows to completely abandon the use
of such representations.

In this paper we restrict ourselves to the simplest possible case of the
$\mathfrak{so}_3$-invariant quantum integrable models.
As already noticed (see end of section \ref{BVsect}), these models are equivalent to the ones built on  the $\mathfrak{gl}_2$ algebra.
However, the calculations leading to the Bethe vectors and the action formulas are rather different: for $\mathfrak{gl}_2$ models, they reflect the general $\mathfrak{gl}_N$ scheme, while $\mathfrak{so}_3$ models are closer to the $\mathfrak{so}_N$ framework.
In this sense, although the calculations presented in this article do not bring any new result on scalar products of Bethe vectors, they shed some light on the case of integrable models based on orthogonal and symplectic Yangians.
Indeed, it is clear that the method introduced here can be generalized to  the
$\mathfrak{so}_N$ and  $\mathfrak{sp}_{2n}$-invariant models. The corresponding results will be published elsewhere.

\section*{Acknowledgments}
The work was performed at the Steklov Mathematical Institute of Russian Academy of Sciences, Moscow.
This work is supported by the Russian Science Foundation under grant 19-11-00062.

\appendix


\begin{thebibliography}{99}

%
\bibitem{FadST79} L. D. Faddeev, E. K. Sklyanin and L. A. Takhtajan, \textsl{Quantum Inverse Problem. I},
 Theor. Math. Phys. {\bf 40} (1979) 688--706.
 %

\bibitem{FadT79} L. D. Faddeev and L. A. Takhtajan, \textsl{The quantum method of the inverse problem and the Heisenberg $XYZ$ model},
Usp. Math. Nauk {\bf 34} (1979) 13--63;  Russian Math. Surveys {\bf 34} (1979) 11--68 (Engl. transl.).


\bibitem{R83} N.~Yu.~Reshetikhin, \textsl{A method of functional equations in the theory of exactly solvable quantum systems},
Lett. Math. Phys. {\bf 7} (1983) 205--213.

\bibitem{R84} N.~Yu.~Reshetikhin, \textsl{$O(N)$ invariant quantum field theoretical models: exact solutions},
Nucl. Phys. B {\bf 251} [FS13] (1985) 565--580.

\bibitem{KulRes81}
P. P. Kulish, N. Yu. Reshetikhin,
\textsl{Generalized Heisenberg ferromagnet and the Gross--Neveu model}, Zh. Eksp. Theor. Fiz.
{\bf 80} (1981) 214--228; Sov. Phys. JETP,  {\bf 53}:1 (1981)  108--114 (Engl. transl.)

 \bibitem{KulRes82}
P. P. Kulish, N. Yu. Reshetikhin,
\textsl{GL(3)-invariant solutions of the Yang-Baxter equation and associated quantum systems},
Zap. Nauchn. Sem. POMI. {\bf 120} (1982) 92--121; J. Sov. Math.,  {\bf 34}:5 (1982)  1948--1971 (Engl. transl.)

\bibitem{KulRes83}
P. P. Kulish, N. Yu. Reshetikhin,
\textsl{Diagonalization of $GL(N)$ invariant transfer matrices and quantum $N$-wave system (Lee model)},
J.~Phys.~A:  {\bf 16} (1983) L591--L596.


\bibitem{VT94}
V. Tarasov, A. Varchenko, \textsl{Jackson inte\-gral re\-pre\-sen\-ta\-ti\-ons of so\-lu\-ti\-ons
of the quan\-tized Knizh\-nik--Za\-mo\-lod\-chi\-kov equation},  Algebra and Analysis, {\bf 6}:2 (1994) 90--137;
St. Petersburg Math. J. {\bf 6}:2 (1995) 275--313 (Engl. transl.), \texttt{arXiv:hep-th/9311040}.

\bibitem{VTcom13}
V. Tarasov, A. Varchenko,
\textsl{Combinatorial formulae for nested Bethe vectors}, SIGMA \textbf{9} (2013) 048,
{\tt arXiv:math/0702277 [math.QA]}.

\bibitem{RS90} N.~Y.~Reshetikhin,   M.~A.~Semenov-Tian-Shansky, \textsl{Central extensions of quantum current groups}
Lett. Math. Phys.  {\bf 19} (1990) 133--142.

\bibitem{JLM18} N.~Jing, M.~Liu, A.~Molev. \textsl{Isomorphism between the $R$-matrix and Drinfeld presentations
of Yangian in types $B$, $C$ and $D$}, Comm. Math. Phys. {\bf 361} (2018) 827--872.

\bibitem{JLY18}  N.~Jing, M.~Liu, F.~Yang. \textsl{Double Yangians of the classical types and their vertex representations},
\texttt{arXiv:1810.06484 [math.QA]}.


\bibitem{HLPRS17} A.~Hutsalyuk,  A.~Liashyk, S.~Z.~Pakuliak, E.~Ragoucy, N.~A.~Slavnov,
\textsl{Current presentation for the double super-Yangian $DY(\mathfrak{gl}(m|n))$ and Bethe vectors},
Russ. Math. Surv. {\bf 72}:1  (2017) 33--99, \texttt{arXiv:1611.09020}.

\bibitem{ZZ79} A.~B.~Zamolodchikov, Al~.B.~Zamolodchikov. \textsl{Factorized $S$-matrices in two dimensions as the exact
solutions of certain relativistic quantum field models}, Ann. Phys. {\bf 120} (1979) 253--291.


\bibitem{KhoPak05}
S. Pakuliak, S. Khoroshkin, \textsl{The weight function for the quantum af\/f\/ine algebra $U_q(\widehat{\mathfrak{sl}}_3)$},
Theor. Math. Phys. {\bf 145} (2005) 1373, \texttt{arXiv:math/0610433 [math-qa]}.

\bibitem{KhP08} S.~Khoroshkin, S.~Pakuliak, \textsl{A computation of an universal weight function for
the quantum affine algebra $U_q(\mathfrak{gl}(N))$},  {J. of Mathematics of Kyoto University},
{\bf 48} n.2 (2008) 277--321, \texttt{arXiv:0711.2819 [math.QA]}.


\bibitem{BPRS12} S.~Belliard, S.~Z.~Pakuliak, E.~Ragoucy, N.~A.~Slavnov, \textsl{Highest coefficient of scalar products in
$SU(3)$-invariant integrable models},  J. Stat. Mech. 1209 (2012) P09003;
\texttt{arXiv:1206.4931 [math-ph]}.

\bibitem{HutLPRS17} A.~Hutsalyuk,  A.~Liashyk, S.~Z.~Pakuliak, E.~Ragoucy, N.~A.~Slavnov,
\textsl{Scalar products of Bethe vectors in the models with $\mathfrak{gl}(m|n))$ symmetry},
Nucl. Phys. B, {\bf 923} (2017) 277--311,\texttt{arXiv:1704.08173}.


\bibitem{M07} A.~Molev, \textsl{Yangian and classical Lie algebras}, Mathematical Surveys and Monographs, 143. Americal
Mathematical Society, Providence, RI, 2007.

\bibitem{AMR06} D. Arnaudon, A.~Molev, E.~Ragoucy, \textsl{On the $R$-matrix realization of Yangians
and their representations}, Annales H. Poincar{\'e} \textbf{7} (2006) 1269, \texttt{math.QA/0511481}.


\bibitem{D88} V.~G.~Drinfeld. \textsl{A new realization of Yangians and of quantum affine algebras}, Soviet Math. Dokl. {\bf 36}
(1988) 212--216.

\bibitem{EKhP07} B.~Enriquez, S.~Khoroshkin, S.~Pakuliak, \textsl{Weight
functions and Drinfeld currents,}
{Comm. Math. Phys.} {\bf 276} (2007) 691--725.

\bibitem{LPRS19} A.~Liashyk, S.~Z.~Pakuliak, E.~Ragoucy, N.~A.~Slavnov, \textsl{New symmetries of $\mathfrak{gl}(N)$-invariant
Bethe vectors}, J. Stat. Mech.: Theory and Experiment (2019) 044001.

\bibitem{Dr88} V.~G.~Drinfeld. \textsl{Quantum groups},
J. Soviet Math., {\bf 41}:2 (1988) 898--915.

\bibitem{KhT96} S.~M.~Khoroshkin, V.~N.~Tolstoy. \textsl{Yangian double}, Lett. Math. Phys. {\bf 36} (1996) 373--402.

\bibitem{KhP05} S.~M.~Khoroshkin, S.~Z.~Pakuliak, \textsl{Weight function for the quantum affine algebra
$U_q(\widehat{\mathfrak{sl}}_3)$}, Theor. Math. Phys. {\bf 145}(1): (2005) 1373--1399.



\bibitem{MukTV} E. Mukhin, V. Tarasov, A. Varchenko, \textsl{Bethe eigenvectors of higher transfer matrices}, J. Stat. Mech. Theory Exp.
{\bf 0608} (2006) P08002, \texttt{arXiv:math/0605015}.
\end{thebibliography}
\end{document}